\documentclass[apjl]{emulateapj}
\usepackage{apjfonts}

\usepackage{amsmath}
\usepackage{amssymb}
\usepackage{epsfig}
\usepackage{natbib}

\newcommand{\kms}{\,{\rm km\,s^{-1}}}
\newcommand{\au}{\,{\rm AU}}

\newcommand{\yr}{\,{\rm yr}}

\newcommand{\gyr}{\,{\rm Gyr}}

\newcommand{\pc}{\,{\rm pc}}

\newcommand{\hr}{\,{\rm hr}}

\newcommand{\msun}{\,M_\odot}
\newcommand{\rsun}{\,R_\odot}

\newcommand{\be}{\begin{equation}}
\newcommand{\ee}{\end{equation}}

\newcommand{\bea}{\begin{eqnarray}}
\newcommand{\eea}{\end{eqnarray}}

\newcommand{\ben}{\begin{enumerate}}
\newcommand{\een}{\end{enumerate}}

\newcommand{\calb}{\mathcal{B}}
\newcommand{\calr}{\mathcal{R}}

\begin{document}

\shorttitle{Relativistic Binary Pulsars}
\shortauthors{Pfahl, Podsiadlowski, \& Rappaport}


\submitted{Submitted to The Astrophysical Journal}

\title{Relativistic Binary Pulsars with Black-Hole Companions}

\author{Eric Pfahl\altaffilmark{1}, Philipp Podsiadlowski\altaffilmark{2}, 
\& Saul Rappaport\altaffilmark{3}}

\altaffiltext{1}{Chandra Fellow; Department of Astronomy, University
of Virginia, 530 McCormick Rd., Charlottesville, VA 22903;
epfahl@virginia.edu}

\altaffiltext{2}{Department of Astrophysics, University of Oxford,
Oxford, OX1 3RH, UK; podsi@astro.ox.ac.uk}

\altaffiltext{3}{Department of Physics, Massachusetts Institute of
Technology, 77 Massachusetts Ave., Cambridge, MA 02139; sar@mit.edu}


\begin{abstract}

Binaries containing a stellar-mass black hole and a recycled radio
pulsar have so far eluded detection.  We present a focused
investigation of the formation and evolution of these systems in the
Galactic disk, highlighting the factors that limit their numbers and
the reasons why they may be extremely rare.  We surmise that the
birthrate of black-hole/recycled-pulsar binaries in the Galactic disk
is probably no higher than $\sim$$10^{-7}\yr^{-1}$, and may be much
less, including zero.  Simple arguments regarding common-envelope
evolution suggest that these binaries should have orbital periods less
than 10\,hr and an average lifetime of $\la$$10^8\yr$ before
coalescence due to the emission of gravitational radiation.  We expect
that fewer than $\sim$10 of these compact, relativistic binaries
currently reside in the Galactic disk, less than 0.1--1\% of the
number of double neutron stars.  The discovery of two or more
black-hole/recycled-pulsar binaries using current radio telescopes
would tightly constrain certain ideas regarding the evolution of
massive stars, dynamical mass transfer, and black-hole formation.

\end{abstract}


\keywords{binaries: close --- black hole physics --- gravitational
  waves --- pulsars: general --- stars: neutron}


\section{INTRODUCTION}\label{sec:intro}

Thirty years have passed since the discovery of PSR B1913+16
\citep{hulse75}, which has the dual distinction of being the first
known binary radio pulsar and double neutron star.  The intervening
years---particularly the last decade---have seen enormous growth in
the observed number and variety of binary radio pulsars. Among the
$\sim$100 known systems (see Stairs 2004 for a representative listing
and references), orbital periods range from 90\,min to several years,
most have white-dwarf companions with masses from substellar to near
the Chandrasekhar limit, three have a stellar companion, and eight
probably contain a second neutron star \citep{stairs04,faulkner05}.

With a handful of exceptions, most radio pulsars in binary systems are
{\em recycled}, characterized by low inferred surface magnetic field
strengths of $\la$$10^{10}\,{\rm G}$ and spin periods of 2--100\,ms.
The term ``recycled'' refers to the accretion of matter and angular
momentum by the neutron star, whereby the spin period is reduced to
small values \citep{smarr76,alpar82,joss83}.  In contrast, the
majority of isolated pulsars (hereafter called ``normal'') have
magnetic fields of $10^{11}$--$10^{13}\,{\rm G}$ and spin periods of
0.1--1\,s.  Recycled pulsars show much greater stability as clocks,
allowing for very precise dynamical measurements if the pulsar is in a
binary.  Precision timing of a recycled binary pulsar quickly reveals
its orbital motion, and if the orbit is sufficiently compact or
inclined, relativistic gravity may cause significant deviations from a
Keplerian model, as in the classic example of PSR B1913+16.

The eight observed double neutron stars (DNSs) have a special status
among binary pulsars.  As exotic endpoints of massive binary stellar
evolution, they probe important details of stellar physics.  Moreover,
since the bright pulsar is recycled, and both components are
essentially point masses, a subset of these systems have proved to be
extremely rich laboratories for studying post-Newtonian gravity.  The
newly discovered {\em double pulsar} binary, PSR J0737--3039, stands
out as having a nearly edge-on orbit and an orbital period of 2.4\,hr,
the shortest among the DNSs \citep{burgay03,lyne04}.  Finally, the
known DNSs, especially PSR J0737--3039, imply a large number of
similar systems in the Milky Way and distant galaxies that will
produce in the next $\sim$1\,Gyr a powerful burst of electromagnetic
and gravitational radiation in the final throes of coalescence
\citep[e.g.,][]{kalogera04}.

All the factors that make DNSs exceptional are greatly enhanced if the
pulsar's companion is replaced with a canonical $10\msun$ black hole.
To date, there has been no detection of a binary containing a
stellar-mass black hole and a recycled pulsar (hereafter BHRP).
Several earlier theoretical studies provide insight into the formation
of BHRPs in the Galactic disk.  \citet{narayan91} first mentioned
these systems, identified a likely formation scenario (see
\S~\ref{sec:prev}), and suggested a birthrate of $\la$$10^{-6}\yr$
based on their nondetection.  Binary population synthesis studies by
\citet{lipunov94}, \citet{sipior02}, \citet{voss03}, and
\citet{sipior04} theoretically constrain the BHRP birthrate and the
expected orbital parameters.  Each of these papers also addresses the
formation of systems where the radio pulsar is not recycled \citep[see
also][]{belkalbul02}, which are not considered here.  These latter
systems are certainly interesting, and probably much more numerous
than BHRPs, although they they are not nearly as desirable as
astrophysical laboratories.  We also do not investigate the formation
of BHRPs in globular clusters \citep{sigurdsson03}.

Our main goals are to (i) dissect the most important steps and
uncertainties in BHRP formation (\S\S~\ref{sec:prev} and
\ref{sec:formation}), (ii) constrain the BHRP birthrate in the
Galactic disk (\S~\ref{sec:birth}), (iii) indicate the range of binary
parameters for newly formed systems and the current-epoch population
(\S~\ref{sec:pop}), and (iv) evaluate the likelihood of discovering
BHRPs in pulsar surveys (\S~\ref{sec:detect}).  Each of these topics
has received some attention in one or more of the above papers, but no
clear picture has yet emerged, and previous work has not examined all
the major physical uncertainties.  We utilize simple, semi-analytic
calculations to present a complete portrait of BHRP formation.  The
next section gives a preview of the major evolutionary hurdles in BHRP
formation, and is followed by a more detailed and quantitative
discussion in \S~\ref{sec:formation}.


\section{Overview of Formation and Uncertainties}\label{sec:prev}

In this paper, we consider the most obvious formation channel for
BHRPs in the Galactic disk, which is qualitatively identical to the
standard model for forming the known DNSs
\citep[e.g.,][]{bhat91,tauris03}\footnote{The only alternative
scenario for DNS formation, involving the spiral-in of two stellar
cores inside a common envelope (Brown (1995), is probably not a viable
channel for BHRP formation, since the pulsar would be the second-born
compact object and hence not be recycled (see also J. Dewi,
Ph. Podsiadlowski, \& A. Sena 2005, in preparation).}.  The scenario
begins with a massive primordial binary, where the primary star is the
neutron-star progenitor of mass $\simeq$$8\msun$ to $\ga$$25\msun$.
During a phase of dynamically stable mass transfer from the primary,
the initially less massive secondary reaches a mass of $\ga$$25\msun$,
and will ultimately leave a black-hole remnant.  Mass transfer ends
when the core of the primary is exposed.  The core then explodes as a
Type Ib/c supernova, leaving behind a neutron star that may have
received a large impulsive kick.  Before the secondary evolves
significantly, the neutron star may be a radio pulsar or an X-ray
source accreting from the wind of its massive companion.  Because of
the extreme binary mass ratio, the second episode of Roche-lobe
overflow is dynamically unstable.  The neutron star is engulfed by the
stellar envelope and dragged toward the secondary's core.  Following
the successful expulsion of the common envelope, the neutron star
orbits the core, which then collapses to a black hole; for less
massive secondaries, a second neutron star is formed at this stage.
Either before, during, or after the spiral-in, the neutron star
accretes a small amount of mass and is recycled.  We now address
several fundamental uncertainties in the evolution of the massive
secondary, spiral-in phase, recycling process, and formation of the
black hole.

\subsection{Stellar Evolution Uncertainties}\label{sec:evunc}

Surprisingly, the evolution of massive stars is still quite poorly
understood.  It is generally estimated that the minimum mass of a {\em
single} star that ends its life as a black hole is $\simeq$$25\msun$
(see Podsiadlowski, Rappaport, \& Han 2003 for discussion and
references; hereafter PRH03).  However, \citet{brown01} show that when
a star less massive than 40--$50\msun$ loses its hydrogen-rich
envelope before helium core burning (case B mass transfer), its core
evolution is changed dramatically, and the star will most likely leave
a neutron-star remnant.  Therefore, mass transfer from the secondary
in the above BHRP formation scenario must begin after it has completed
most or all of its helium core burning phase \citep[case C mass
transfer; see also][]{wellstein99}.  The formation of BHRPs is then
very sensitive to the radial evolution of the massive secondary
\citep[see also][]{zwart97} and uncertainties in the stellar physics.

The range of binary separations for case C mass transfer depends
mainly on the difference between the maximum radius of a single star
and the radius at the tip of the first giant branch. For massive
stars, this range may be as small as a few percent of the radius in
some evolutionary models or a factor of $\ga$2 in others.  The result
depends on when stars undergo helium core burning (see, e.g., the
difference between the evolutionary tracks in Fig.~1 and 2 of Pfahl,
Rappaport, \& Podsiadlowski 2002; hereafter PRP02). If helium burning
occurs only while the star is a blue supergiant, the range of case C
mass transfer is greatly increased compared to when the star burns
helium as a red supergiant (by perhaps a factor of 4 or more; Dewi et
al. 2005, in preparation).  The observed distribution of supergiants
in the Hertzsprung-Russell diagram \citep[e.g.,][]{humphreys84}
suggests that stars as massive as $\sim$$40\msun$ spend a significant
fraction of their helium burning phase as both blue and red
supergiants.  None of the present theoretical models is yet fully
consistent with the observations \citep[e.g.,][]{langer95}.

\subsection{Severity and Implications of the Spiral-In}\label{sec:cesev}

For a neutron star sinking into the envelope of a $\ga$$25\msun$ star,
the orbit may have to shrink by a factor of $\ga$$10^3$ before the
envelope can be expelled, according to recent calculations of envelope
binding energies (Dewi \& Tauris 2001; PRH03).  In order to avoid a
merger of the neutron star and stellar core of radius $\sim$$1\rsun$,
the initial orbital separation must be $\ga$$2000\rsun$.  Stars of
mass 25--$50\msun$ attain maximum radii of $\simeq$1500--$2500\rsun$
\citep[e.g.,][]{hurley00}, and so the range of orbital separations for
successful systems is restricted to $\simeq$2000--$4000\rsun$. A star
that fills its Roche lobe in such an orbit is very evolved, and in
most cases has passed through the phase of helium core burning. Thus,
in the \citet{brown01} picture, we are safe in using $\simeq$$25\msun$
as the mass threshold for black-hole formation.

If the orbits contract by a factor of 1000, the final separations are
$\la$$4\rsun$.  However, if the contraction factors are generally
$\ga$2000, which is quite plausible, then it may be that {\em no BHRPs
are formed}.  Another serious complication arises because the
secondaries of interest are very evolved prior to the spiral-in.
Prodigious stellar winds during the late phases of evolution may
generally cause the secondary's Roche lobe to expand faster than the
star, precluding mass transfer altogether (see \S~\ref{sec:secwind};
Portegies Zwart, Verbunt, \& Ergma 1997; Kalogera \& Webbink 1998;
PRH03).  On the other hand, if the star can catch its Roche lobe after
losing a significant amount of mass, the orbit is wider at the start
of the spiral-in and the envelope mass that must be ejected is
reduced, giving the system a better chance to survive..

We assume that the neutron star does not accrete at hypercritical
rates and collapse to a black hole during the spiral-in
\citep[e.g.,][]{chevalier93,chevalier96}.  If this were to occur
generally, as may, in fact, be the case, then neither DNSs nor BHRPs
would form by the channel being discussed \citep[e.g.,][]{brown95}.

\subsection{Recycling}\label{sec:recycling}

The neutron star in a BHRP is mildly recycled after accreting a small
amount of matter ($\la$$0.001\msun$) from, e.g., the wind of its
companion prior to the spiral-in, the stellar envelope during the
spiral-in itself, or a residual accretion disk after the orbital
contraction has halted.  For most DNSs, recycling may result during a
short phase of stable mass transfer from the Roche-lobe filling core
of the secondary if it is sufficiently low in mass and if it is
exposed before central helium ignition \citep{dewi03,ivanova03}.
However, this process does not apply to BHRPs, because secondary's
core is too massive for the transfer to be stable.

An interesting possibility is if recycling in BHRPs occurs mainly
after the spiral-in, as the neutron star accretes a fraction of the
strong wind from the secondary's core.  Since the core has most likely
passed through the phase of central helium burning, only
$\la$$10^4\yr$ remains before it collapses.  At the Eddington-limited
accretion rate of $\simeq$$4\times 10^{-8}\msun\yr^{-1}$ (for pure
helium), the neutron star can accept only $\sim$$10^{-4}\msun$ of
material.  The short timescale, low accreted mass, and possibly low
angular momentum of material accreted from a wind, all suggest that
pulsars in BHRPs may not be strongly recycled, and could have
systematically longer spin periods and higher surface magnetic fields
than pulsars in DNSs (see Arzoumanian, Cordes, \& Wasserman 1999 for a
related discussion and references).

\subsection{Collapse of the Secondary}\label{sec:collapse}

Uncertainties in black-hole formation include the effects of winds
from the massive progenitor and its exposed core in a binary system,
stellar rotation, and the dynamics of core collapse (e.g., Fryer \&
Kalogera 2001 and references therein).  For the collapse, there are
two possibilities.  Either the core collapses promptly without being
accompanied by a bright supernova \citep[e.g.,][]{gourgoulhon93}, or a
neutron star forms first and subsequently collapses to a black hole
after accreting fallback material \citep{woosley95}, ejecting a
significant fraction of the envelope and resulting in a successful
supernova.  The two-stage scenario has been inferred for the
black-hole progenitor in the binary Nova Scorpii \citep[GRO
J1655--40;][]{podsi02}.  Moreover, the high space velocity of Nova Sco
requires that the black hole received a large natal kick of
$\ga$$100\kms$ (Brandt, Podsiadlowski, \& Sigurdsson 1995). It is not
known if most black holes receive significant kicks at birth
\citep[e.g.,][]{fryer01,jonker04}.  A plausible guess is that the kick
{\em momentum} received by newborn black holes is similar to that
received by neutron stars (see \S~\ref{sec:bhform}).


\section{Formation Details}\label{sec:formation}

Here we elaborate on the points in the last section.  Each stage in
the formation of a BHRP is described explicitly and quantitatively,
although with a more transparent approach compared to many previous
binary population synthesis studies.  Our chosen assumptions and
prescriptions for stellar evolution are fairly standard, but are not
unique.  Further details and references may be found in PRP02.

\subsection{Primordial Binaries}\label{sec:pb}

In our standard model, the mass of the primary is drawn from a
power-law initial mass function, $p(M_p^{\rm PB}) \propto (M_p^{\rm
PB})^{-2.5}$, where $M_p^{\rm PB}$ is in the range of $8\msun$ to
$M_{\rm th}$, the mass threshold for black-hole formation.  Because of
the uncertainty in $M_{\rm th}$ and its possible dependence on binary
evolution (see \S~\ref{sec:evunc}), we consider two scenarios, one
where $M_{\rm th} = 25\msun$ (scenario I), and the other where $M_{\rm
th} = 40\msun$ (scenario II).  The secondary mass $M_s^{\rm PB}$ is
taken from a flat distribution in mass ratios, $q_{\rm PB} = M_s^{\rm
PB}/M_p^{\rm PB} <1$.  Alternative choices for $p(M_p^{\rm PB})$ and
$p(q_{\rm PB})$ are considered in \S~\ref{sec:pop} and Table~1.
Without much loss in generality, we assume that the orbit is circular,
with a separation, $a_{\rm PB}$, drawn from a logarithmically flat
distribution that extends from $\sim$$10\rsun$ to a somewhat arbitrary
maximum of $10^6\rsun$.

\begin{figure}
\centerline{\epsfig{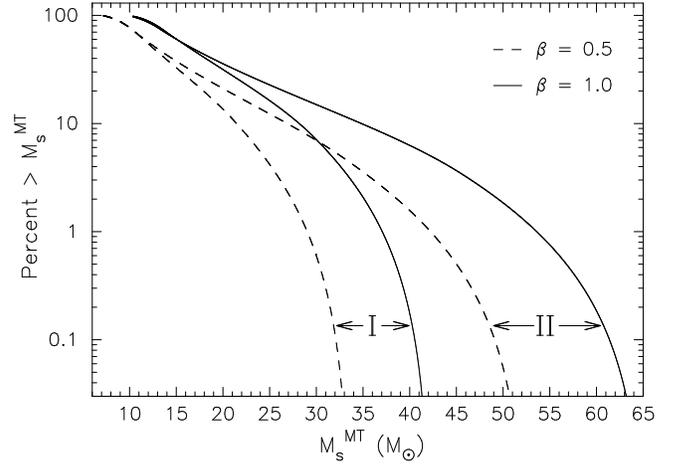}}
\caption{Cumulative distribution of secondary masses following stable
mass transfer, where the primary initial mass function is $p(M_p^{\rm
PB}) \propto (M_p^{\rm PB})^{-2.5}$ for $8\msun < M_p^{\rm PB} <
M_{\rm th}$, and the secondary mass is drawn from a flat distribution
in mass ratios such that $M_s^{\rm PB}/M_p^{\rm PB} > 0.5$.  The mass
capture fraction is fixed at $\beta = 0.5$ (dashed curves) or 1.0
(solid curves).  Each pair of curves for a given $\beta$ corresponds
to a different mass threshold for black-hole formation from the {\em
primary}: $M_{\rm th} = 25\msun$ (scenario I) or $M_{\rm th} =
40\msun$ (scenario II).
\label{fig:seccum}}
\end{figure}

\subsection{Stable Mass Transfer}\label{sec:stabmt}

In isolation, the primary would grow to a maximum radius of
$\ga$$1000\rsun$ before exploding as a supernova.  Mass transfer from
the primary occurs if it fills its Roche lobe of radius $R_{\rm L} =
a_{\rm PB} r_{\rm L}(q_{\rm PB})$, where \citep{eggleton83}
\begin{equation}\label{eq:roche}
r_{\rm L} = \frac{0.49}{0.6+q_{\rm PB}^{2/3}\ln(1+q_{\rm PB}^{-1/3})} ~,
\end{equation}   
with typical values of $r_{\rm L} = 0.5\pm 0.1$.  Mass transfer occurs
over $\simeq$2 decades in $a_{\rm PB}$, comprising $\simeq$40\% of the
binaries.

Mass transfer may be stable or dynamically unstable depending on
$q_{\rm PB}$ and the evolutionary state of the primary.  If mass
transfer is to be stable, then $q_{\rm PB}$ cannot be too small, and
the primary should not be a deeply convective red supergiant (see
PRP02 for a discussion, caveats, and references).  In this regard, we
assume that $q_{\rm PB} > 0.5$, which is true for half of the
binaries.  By excluding very evolved primaries at the onset of mass
transfer, we limit the stellar radius to less than
$\simeq$200--$1000\rsun$, where larger values correspond to higher
masses.  We also neglect systems where the primary transfers matter on
the main sequence ($a_{\rm PB} \la 30$--$40\rsun$), which will not
evolve into the objects of interest here.  With these cuts, the net
fraction of massive binaries that undergo stable mass transfer after
the primary has left the main sequence is $f_{\rm MT} \sim 0.1$.

The primary is expected to lose its entire envelope during mass
transfer, leaving its hydrogen-exhausted core of mass
\citep[e.g.,][]{hurley00}
\begin{equation}\label{eq:coremass}
M_{c,p} \simeq 0.1(M_p^{\rm PB}/M_\odot)^{1.35}\msun~.
\end{equation}
Let $\beta$ be the fraction of transferred matter accreted by the
secondary.  The final secondary mass is then
\begin{equation}\label{eq:secmass}
M_s^{\rm MT} = M_s^{\rm PB} + \beta(M_p^{\rm PB} - M_{c,p})~.
\end{equation}
We assume that the secondary is reset on the main sequence appropriate
for its new mass, which is not justified for $q_{\rm PB}$ near 1.

Given $\beta$, it is straightforward to determine the distribution of
$M_s^{\rm MT}$ given the distributions of $q_{\rm PB}$ and $M_p^{\rm
PB}$.  Figure~\ref{fig:seccum} shows the resulting cumulative
distribution of $M_s^{\rm MT}$ for $\beta = 0.5$ and 1.0 in scenarios
I and II.  The fraction of systems in which the secondary collapses to
a black hole and that remain potential BHRP progenitors can be
estimated from Fig.~\ref{fig:seccum}.  If, in accord with the remarks
in \S~\ref{sec:cesev}, we assume that all post-transfer secondaries of
mass $>$$25\msun$ leave black-hole remnants, then for $\beta =
0.5$--1, the fraction of systems in scenario I is $f_{\rm BH} \simeq
0.04$--0.16, while in scenario II the fraction is $f_{\rm BH} \simeq
0.12$--0.23.  Setting the minimum secondary mass to $20\msun$, we find
that $f_{\rm BH}$ spans the range of 0.2--0.37 and 0.13--0.3 in
scenarios I and II, respectively.  Likewise, a minimum mass of
$30\msun$ gives fractions of 0.07--0.14 and 0.006--0.14 in the two
scenarios.

The orbital separation following stable mass transfer, $a_{\rm MT}$,
depends sensitively on $\beta$ and the specific orbital angular
momentum taken by escaping matter.  A plausible first approximation is
that $a_{\rm MT} \sim a_{\rm PB}$.  The resulting $\log a_{\rm MT}$
distribution is then roughly flat from $a_{\rm MT} \sim 0.1\au$ to
several AU, with a declining tail that extends to $\simeq$10\,AU,
which is correlated with higher post-transfer secondary masses.  For
separations $\ga$10\,AU, the primary cannot fill its Roche lobe during
the mass transfer phase.  

\begin{figure}
\centerline{\epsfig{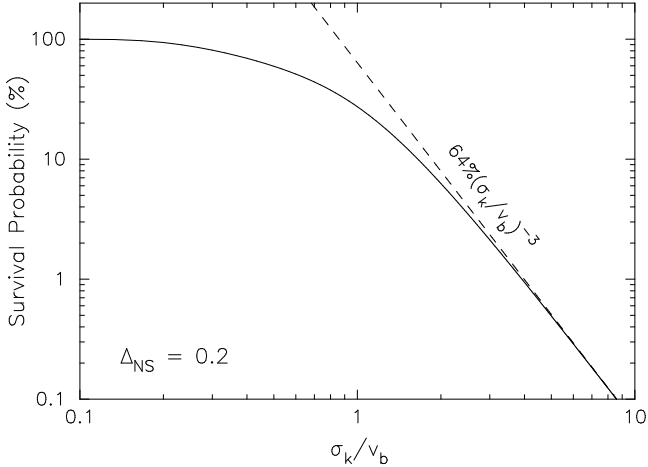}}
\caption{Supernova survival probability as a function of
$\sigma_k/v_b$ for a Maxwellian kick-speed distribution, where
$\sigma_k$ is the one-dimensional velocity dispersion.  The
probability is $<$10\% for $\sigma_k/v_b \ga 1.7$ and $<$1\% for
$\sigma_k/v_b \ga 3.9$. For $\sigma_k/v_b \ga 2$, the probability
approaches the cubic trend $\simeq$$64\% (\sigma_k/v_b)^{-3}$ (dashed
line). \label{fig:survint}}
\end{figure}

\subsection{Supernova of the Primary}\label{sec:postsn}

After mass transfer, $\la$1\,Myr remains before the Type Ib/c
supernova explosion of the primary's exposed core and formation of the
neutron star, for which we assume a mass of $M_{\rm NS} = 1.4\msun$.
Radial expansion of the naked core of the primary and wind mass loss
do not affect our main results and are neglected.  Associated with the
core collapse and supernova is impulsive mass loss and a kick to the
newborn neutron star.  Measured proper motions of isolated radio
pulsars suggest a median kick speed of 100--$300\kms$
\citep[e.g.,][]{hansen97,arzoumanian02}.  The functional form of the
underlying kick distribution is weakly constrained.  Moreover, there
are observational and theoretical reasons to believe that the average
kick magnitude may be much smaller for many neutron stars born in
binary systems \citep{pfahl02b,podsi04}.  We now quantify the
dynamical effects of supernova mass loss and kicks on potential BHRP
progenitors with $M_s^{\rm MT} > 25\msun$.

Since the primary's core mass is $\la$$8\msun$, for BHRP progenitors
the fractional binary mass lost in the supernova is
\begin{equation}
\Delta_{\rm NS} = \frac{M_{c,p} - M_{\rm NS}}{M_{c,p} + M_s^{\rm MT}} 
< 0.2~,
\end{equation}
and so the dynamical effect of mass loss is relatively small.  There
is a significant probability for a binary to be unbound by the
supernova if the neutron-star kick speed, $v_k$, is comparable to or
larger than the relative orbital speed of the binary components before
the explosion:
\begin{equation}
v_b 
\simeq 190\kms\left(\frac{M_b^{\rm MT}}{40\msun}
\cdot\frac{1\au}{a_{\rm MT}}\right)^{1/2}~,
\end{equation}
where $M_b^{\rm MT} = M_s^{\rm MT} + M_{c,p}$.  When $v_k/v_b = 1$,
the probability that the binary remains intact is roughly 50\% if the
kick directions are distributed isotropically.  To be more
quantitative, we must adopt a specific $v_k$ distribution.

A Maxwellian distribution, $p(v_k) \propto v_k^2
\exp(-v_k^2/2\sigma_k^2)$, has the attractive features of being a
reasonable characterization of the pulsar velocity data (for $\sigma_k
= 100$--$200\kms$) and having only one free parameter.
Figure~\ref{fig:survint}, the result of a simple numerical
integration, shows the survival probability versus $\sigma_k/v_b$ for
an assumed fractional mass loss $\Delta_{\rm NS} = 0.2$.  In
Fig.~\ref{fig:survint}, the probability asymptotically approaches
$\simeq$$0.64(\sigma_k/v_b)^{-3}$, a trend which is a direct
consequence of the Maxwellian distribution.  When $\sigma_k = 100
\kms$ ($200\kms$), the survival probability drops below 10\% when
$a_{\rm MT} \ga 2.5\au$ (0.6\,AU) for a reference binary mass of
$40\msun$.

\begin{figure}
\centerline{\epsfig{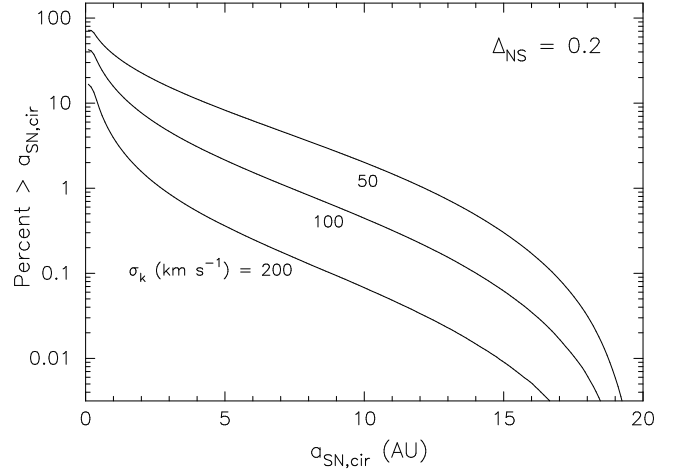}}
\caption{Cumulative distribution of circularized, post-supernova
orbital separations, calculated under the assumptions that the
pre-supernova binary mass is $40\msun$, 20\% of the binary mass is
lost in the explosion, and the distribution of pre-supernova orbital
separations is logarithmically flat from 0.2\,AU to 10\,AU.
\label{fig:circ}}
\end{figure}

More input is needed to estimate the net fraction of binaries that
remain viable BHRP progenitors after the supernova.  The semimajor
axis and eccentricity after the supernova are given by
\citep[e.g.,][]{brandt95}
\begin{equation}\label{eq:asn}
\left(\frac{a_{\rm SN}}{a_{\rm MT}}\right)^{-1} = 1 - 
\frac{\Delta_{\rm NS} + 2\xi w + w^2}{1 - \Delta_{\rm NS}}~,
\end{equation}
and
\begin{equation}\label{eq:esn}
1 - e_{\rm SN}^2 = 
\left(\frac{a_{\rm SN}}{a_{\rm MT}}\right)^{-1}
\frac{(1 + \xi w)^2 + \zeta^2 w^2}{1-\Delta_{\rm NS}}~,
\end{equation}
where $w = v_k/v_b$, and $v_k\xi$ and $v_k\zeta$ are the projections
of the kick velocity onto, respectively, the directions of the
pre-supernova orbital velocity and orbital angular momentum.  After
the supernova, the orbit circularizes to a separation of $a_{\rm
SN,cir} = a_{\rm SN}(1 - e_{\rm SN}^2)$---assuming conservation of
orbital angular momentum---before the secondary fills its Roche lobe.
At this stage, we desire the distribution of $a_{\rm SN,cir}$ for
surviving binaries ($e_{\rm SN} < 1$).  To this end, we have adopted
the Maxwellian kick distribution, assumed for illustrative purposes
that the $a_{\rm MT}$ distribution is logarithmically flat from
0.2\,AU to 10\,AU, and fixed the total binary mass at $40\msun$.  For
reasons discussed in \S~\ref{sec:stabmt}, this choice for the $a_{\rm
MT}$ distribution may significantly overestimate the number of wide
binaries.  The elementary constraint $a_{\rm SN}(1 - e_{\rm SN}) <
a_{\rm MT}$ implies that $a_{\rm SN,cir} < 2a_{\rm MT}$ for $e_{\rm
SN} < 1$; thus, $a_{\rm SN,cir} < 20\au$ in this example.
Figure~\ref{fig:circ}, the result of a simple Monte Carlo integration,
shows the cumulative distribution of $a_{\rm SN,cir}$ for systems that
remain bound.  For $\sigma_k = 50\kms$, $100\kms$, and $200\kms$, the
fraction of binaries with $a_{\rm SN,cir} > 10\au$ is $\simeq$2\%
$\simeq$0.4\%, and $\simeq$0.07\%, respecitively.  Most systems with
$a_{\rm SN,cir} < 10\au$ ($\simeq$$2000\rsun$) will not survive the
common-envelope phase, as discussed in \S~\ref{sec:cesev}.  The
cumulative probability scales approximately as $\sigma_k^{-2.5}$ for
$a_{\rm SN,cir} > 5\au$.  An increase in the total binary mass from
$40\msun$ to $60\msun$ increases the probabilities by at most a factor
of 2.  In \S~\ref{sec:cedet}, we combine the results in
Fig.~\ref{fig:circ} with the severe constraints of the spiral-in phase
in order to more precisely estimate the fraction of binaries that
survive both the first supernova and the spiral-in.

Supernova mass loss and the neutron-star kick impart some velocity to
the binary center of mass.  Considered independently, the
contributions to the systemic speed from mass loss and the kick are,
respectively, $v_b \Delta_{\rm NS} (1 + M_{\rm NS}/M_s^{\rm MT})^{-1}$
and $v_k(1 + M_s^{\rm MT}/M_{\rm NS})^{-1}$ (e.g., Appendix B of
PRP02).  Most binaries that survive to ultimately form BHRPs will have
$v_b$ and $v_k$ both less than $100\kms$, since only wide systems can
survive the common-envelope phase.  Therefore, given that $\Delta_{\rm
NS} \la 0.2$ and $M_{\rm NS}/M_s^{\rm MT} < 0.06$, the supernova of
the primary's core imparts systemic speeds that are typically
$<$$10\kms$.

\subsection{Secondary Winds}\label{sec:secwind}

When the secondary's radius is $\ga$$1000\rsun$, it may lose most of
its hydrogen envelope in a wind.  A spherically symmetric wind that
carries the specific orbital angular momentum of the star causes the
orbit to expand at a rate of $\dot{a}/a = \dot{M}_w/M_b$, where
$\dot{M}_w > 0$ is the mass loss rate, and $M_b$ is the instantaneous
binary mass.  During this phase of wind mass loss, the Roche-lobe
radius of the secondary expands approximately in proportion to
$M_b^{-0.85}$.  If the stellar radius expands with a shallower mass
dependence, which is quite possible for the masses of interest, then
it will not be allowed to fill its Roche lobe at this stage (PRH03;
see also Portegies Zwart, Verbunt, \& Ergma 1997 and Kalogera \&
Webbink 1998 for related discussions).  However, stellar winds from
massive stars and the effects on binary evolution are very uncertain.
The prescription for the angular momentum taken by the wind can be
modified so as to limit or reverse the orbital expansion (see PRH03),
making it possible for the secondary to fill its Roche lobe after its
mass has been reduced significantly by the wind.  This may lead to
less extreme shrinkage during the common-envelope phase (see below),
although it seems that the conditions that permit both Roche-lobe
overflow and significant mass loss may require fine tuning.  This
aspect of black-hole binary formation deserves considerable attention.

\subsection{Common-Envelope Phase}\label{sec:cedet}

As the neutron star spirals in to the stellar envelope, frictional
luminosity is generated at the expense of the orbital binding energy.
If sufficient energy is available to disperse the envelope, the
neutron star emerges in a tight orbit about the stellar core.  A
standard energetics argument \citep{webbink84} gives the orbital
separation following the spiral-in:
\begin{equation}\label{eq:ce}
\frac{a_{\rm CE}}{a_{\rm SN,cir}} \simeq
\gamma\frac{r_{\rm L}(q_{\rm SN})}{2} 
\frac{M_{\rm NS} M_{c,s}}{M_s^{\rm MT} M_{e,s}}~, 
\end{equation}
where $q_{\rm SN} = M_{\rm NS}/M_s^{\rm MT} \ll 1$, $M_{c,s}$ and
$M_{e,s}$ are the secondary's core and envelope mass, and $\gamma$ is
a combined parameter that quantifies the binding energy of the
envelope and the efficiency with which orbital energy is used to expel
the envelope.  Over the full range of $M_s^{\rm MT}$, a good
approximation to eq.~(\ref{eq:ce}) is 
\begin{equation}
  \frac{a_{\rm CE}}{a_{\rm SN,cir}} \simeq 0.007\gamma
\end{equation}
if stellar winds are neglected.  Stellar winds prior to Roche-lobe
overflow reduce the envelope mass but not the core mass.  A 50\%
reduction in the envelope mass increases $a_{\rm CE}/a_{\rm SN,cir}$
by a factor of $\simeq$3, allowing a larger number of systems to avoid
merger.  However, as discussed in the last section, a star that
experiences this degree of wind mass loss is not likely to fill its
Roche lobe at all.

The structure parameter $\gamma$ varies as a given star evolves.
Moreover, as the stellar mass is increased, the value of $\gamma$ at a
particular evolutionary stage decreases systematically.  Stars of mass
$>$$25\msun$ are expected to have $\gamma < 0.1$ in the late stages of
evolution (Dewi \& Tauris 2001; PRH03), and as low as $\sim$$0.01$,
giving contraction factors of $\ga$1400 from eq.~(\ref{eq:ce}).  The
details of common-envelope evolution are poorly understood, especially
the hydrodynamical aspects.  Thus, there is considerable uncertainty
in the appropriate $\gamma$, although it is difficult to see how the
orbital contraction factor for the systems of interest could be much
less than 1000 ($\gamma \simeq 0.14$).

\begin{figure}
\centerline{\epsfig{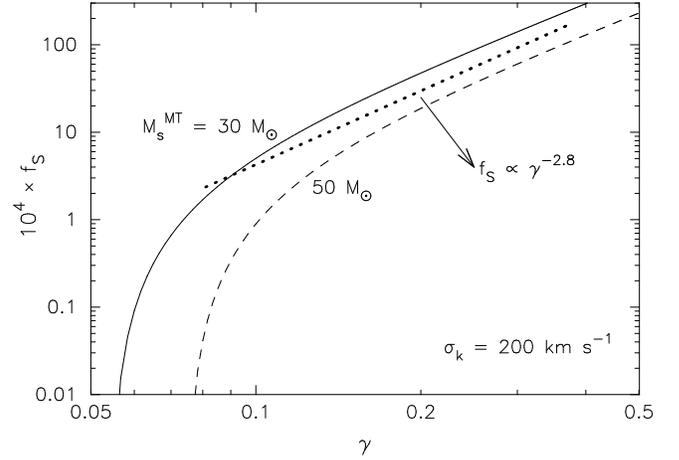}}
\caption{Fraction of systems that survive the both the first supernova
and the spiral-in phase (see text for details), assuming $\sigma_k =
200\kms$.  The two curves correspond to secondary masses of $M_s^{\rm
MT} = 30\msun$ (solid) and $50\msun$ (dashed).  The thick dotted line
shows a power-law approximation intermediate between the two curves
curves for $\gamma \ga 0.15$.
\label{fig:fcegam}}
\end{figure}

We now apply the results of \S~\ref{sec:postsn} and illustrate how the
fraction of systems that survive the spiral-in depends on $\gamma$.
The core radius is estimated to be (J. Dewi \& O. Pols, private
communication)
\begin{equation}\label{eq:corerad}
R_{c,s} = 0.9(M_{c,s}/10\msun)^{0.6}\rsun~.
\end{equation}
We assume a merger results if the core overflows its Roche lobe at the
end of the spiral-in, i.e., if $a_{\rm CE} < R_{c,s}/r_{\rm L}(q_{\rm
CE})$, where $q_{\rm CE} = M_{\rm NS}/M_{c,s}$.  The merger condition
then defines the allowed range in $a_{\rm SN,cir}$:
\begin{equation}
a_{\rm SN,cir} \ga 6.7\,
r_{\rm L}^{-1}(q_{\rm CE})
\left(\frac{\gamma}{0.1}\right)^{-1} 
\left(\frac{R_{c,s}}{R_\odot}\right)
\au~.
\end{equation}
We can now extract from Fig.~\ref{fig:circ} the fraction, $f_{\rm S}$,
of systems that survive both the first supernova explosion and the
common-envelope phase.  Results are shown in Fig.~\ref{fig:fcegam} for
$\sigma_k = 200 \kms$ and secondary masses of $M_s^{\rm MT} = 30\msun$
and $50\msun$.  Smaller values of $f_{\rm S}$ for the $50\msun$ case
are due to the larger radius of the exposed core.  We have not taken
into account the proper mass dependence of the supernova survival
probability.  However, this only partially makes up the difference
between the two curves in Fig.~\ref{fig:fcegam}.  Also shown in
Fig.~\ref{fig:fcegam} is a steep power-law approximation ($f_{\rm S}
\propto \gamma^{2.8}$) to the survival fraction that works well for
$\gamma > 0.15$, but is clearly an overestimate when $\gamma \la 0.1$.
If this `fit' is combined with the $\sigma_k$ scaling from
\S~\ref{sec:postsn}, we arrive at the simple approximate formula,
\begin{equation}\label{eq:fsfit}
f_{\rm S} \sim 5\times 10^{-4}
\left(\frac{\sigma_k}{200\kms}\right)^{-2.5} 
\left(\frac{\gamma}{0.1}\right)^{2.8}~.
\end{equation}

The survival fraction that results from the above procedure assumes
that the star can always grow large enough to fill its Roche lobe.  In
fact, the stars of interest probably grow no larger than
$\simeq$$2500\rsun$, which limits the maximum orbital separation to
$\simeq$$4000\rsun$ or $\simeq$$19\au$.  A smaller maximum radius only
serves to decrease $f_{\rm S}$, exacerbating the difficulty in forming
BHRPs. If the orbit contracts by a factor of $>$1000 during the
spiral-in, the resulting orbital separations are $\la$$4\rsun$, giving
orbital periods of $\la$7\,hr and relative orbital speeds of
$\ga$$700\kms$ for a total binary mass of $10\msun$.

\subsection{Supernova of the Secondary and Final Products}\label{sec:bhform}

The severity of the orbital contraction in the common-envelope phase
implies that the orbit must initially be very wide and the secondary
very evolved in order for the system to avoid merger.  In most, if not
all, cases the exposed core of the secondary will have already
exhausted its central helium supply, so that its remaining lifetime
following the spiral-in is $\la$$10^4\yr$.  At a mass-loss rate of
$10^{-5}$--$10^{-4}\msun\yr^{-1}$ \citep[e.g.,][]{chiosi86}, stellar
winds only have time to take away $\la$$1\msun$, leading to a
$\la$10\% widening of the orbit.

If the eventual core collapse is accompanied by a successful supernova
explosion (see, however, \S~\ref{sec:collapse}), as much as half the
mass of the secondary's core may be ejected, leading to a large
orbital eccentricity.  If there is no kick to the black hole, the
post-supernova eccentricity is $e_{\rm SN} = \Delta_{\rm
BH}/(1-\Delta_{\rm BH})$, where $\Delta_{\rm BH}$ is the fractional
mass loss from the binary (eqs.~[\ref{eq:asn}] and [\ref{eq:esn}]).
Black holes probably do receive kicks at birth, due to asymmetries in
the collapse, explosion, or in the accretion of fallback material.
However, even a black-hole kick speed of $\sim$$100\kms$---as inferred
for Nova Sco \citep{brandt95b}---does not have a dramatic effect on
the orbit, since the pre-supernova orbital speeds are $\ga$$700\kms$.
It is reasonable to suppose that, on average, black holes receive a
kick {\em momentum} that is similar to neutron stars, leading to a
typical kick speed that is perhaps $(100$--$300)\times (M_{\rm
NS}/M_{\rm BH})\kms$, with values of $\sim$10--$80\kms$ for $M_{\rm
BH} = 5$--$10\msun$.  If typical mass-loss and kick parameters are
$\Delta_{\rm BH} \la 0.3$ and $w \la 0.2$, most newly formed BHRPs
will have $e_{\rm SN} \la 0.5$.  Higher eccentricities are correlated
with larger post-supernova semimajor axes, perhaps $\simeq$50\% larger
than the maximum separation following the spiral-in phase.  In
general, we expect that few systems are unbound at the time of
black-hole formation.

After the formation of the black hole, further orbital evolution is
driven by the emission of gravitational radiation. The elapsed time
before gravitational radiation causes the black hole and neutron star
to merge is 
\begin{equation}\label{eq:grtime} 
\tau_m\simeq 2.1\,{\rm Myr} \,
P_h^{8/3} M_{10}^{-2/3}\mu_1^{-1}(1 - e^2)^{7/2}~,
\end{equation} 
where $P_h$ and $e$ are the orbital period (in hours) and
eccentricity, $M_{10} = (M_{\rm NS} + M_{\rm BH})/10\msun$, and
$\mu_1$ is the reduced mass in $M_\odot$.  Equation~(\ref{eq:grtime})
is our own fit (good to better than 20\% for $e < 0.9$) to the full
integral expression for the merger time given in \citet{peters64}.
BHRPs with relatively small eccentricities and periods of 5--10\,hr
merge in $10^8$--$10^9\yr$.

In general, the mass loss and kick that coincide with the formation of
the black hole make the largest contribution to the systemic velocity
of a newly formed BHRP.  Because the black hole dominates the mass of
the binary, most of the kick velocity is transferred to the system; an
average speed might be 20--$30\kms$ if the kick momenta are similar to
neutron stars.  If the average orbital separation following the
spiral-in of the neutron star is 2--$3\rsun$, then the average
pre-supernova orbital speed is 800--$1000\kms$.  Thus, the mean
systemic speed due to impulsive mass loss in the second supernova may
be $\simeq$$20(\Delta_{\rm BH}/0.1)\kms$, where $\Delta_{\rm BH}$ is a
typical value in this context.  A typical systemic speed of a
$\text{few}\times 10\kms$ results in a vertical scaleheight in the
Galactic disk of a $\text{few}\times 100\pc$.  In contrast, it has
been suggested that DNSs are born with mean systemic speeds of
$\ga$$100\kms$, acquired mainly after the second supernova, giving
this population a scaleheight of $\ga$1\,kpc.


\section{Birthrate Estimate}\label{sec:birth}

We take the total formation rate of massive binaries to be the
core-collapse supernovae rate, $\calr_{\rm SN}$
\citep[$\sim$$10^{-2}\yr^{-1}$;][]{cappellaro99}, multiplied by the
fraction, $f_b \simeq 0.5$, of stars in binaries.  Thus,
$f_b\calr_{\rm SN}$ closely approximates the formation rate of the
primordial binaries discussed in \S~\ref{sec:pb}.  Collected and
summarized below are the efficiencies with which the initial ensemble
of binaries passes through the various stages of evolution described
above:

\medskip

\noindent
$f_{\rm MT}$: The fraction of massive primordial binaries that undergo
stable mass transfer after the primary has left the main sequence (see
\S\S~\ref{sec:pb} and \ref{sec:stabmt} for details).  For a
logarithmically flat distribution in initial orbital separations,
$\simeq$10\% of the massive binaries satisfy our criteria for stable
mass transfer.  Different assumptions that do not depart drastically
from ours may lead to variations in $f_{\rm MT}$ by a factor of 2 or
3.

\medskip

\noindent
$f_{\rm BH}$: The fraction of post-transfer secondaries that
ultimately leave black-hole remnants.  In addition to assumptions
about the distributions of masses and orbital separations for the
primordial binaries, this fraction depends on the mass threshold for
black-hole formation and on the amount of matter the secondary is able
to accept from the primary.  Figure~\ref{fig:seccum} suggests that
$f_{\rm BH} \simeq 0.04$--0.2.

\medskip

\noindent
$f_{\rm S}$: The fraction of systems that survive the supernova of the
primary and avoid a merger following the spiral-in phase.
Figure~\ref{fig:fcegam} shows the strong sensitivity of $f_{\rm S}$ to
$\gamma$ in the specific case of $\sigma_k = 200\kms$. An approximate
scaling of $f_{\rm S}$ with $\sigma_k$ and $\gamma$ is given in
eq.~(\ref{eq:fsfit}), which overestimates $f_{\rm S}$ for $\gamma \la
0.1$.

\medskip

\noindent
Combining these factors, we estimate the Galactic birthrate of BHRPs:
\begin{equation}\label{eq:rate}
\calb 
 =  5\times 10^{-8}\yr^{-1} 
\frac{\calr}{0.01\yr^{-1}} \cdot
\frac{f_b}{0.5} \cdot
\frac{f_{\rm MT}}{0.1} \cdot
\frac{f_{\rm BH}}{0.1} \cdot
\frac{f_{\rm S}}{0.001}~.
\end{equation}
Both $f_{\rm MT}$ and $f_{\rm BH}$ could be made larger by a factor of
2 or 3 over their fiducial values in eq.~(\ref{eq:rate}).  It is then
conceivable that, if both $f_{\rm MT}$ and $f_{\rm BH}$ are increased
significantly, $\calb$ could be as large as $\sim$$10^{-6}\yr^{-1}$
even if $f_{\rm S} \sim 10^{-3}$.  However, it is important to note
that $f_{\rm S} \sim 10^{-3}$, which corresponds to $\gamma \ga 0.1$,
is quite optimistic in light of recent envelope binding energy
calculations. We claim that a plausible upper limit to the rate is
$\sim$$10^{-7}\yr^{-1}$.


%
\begin{figure}[t]
\centerline{\epsfig{file = 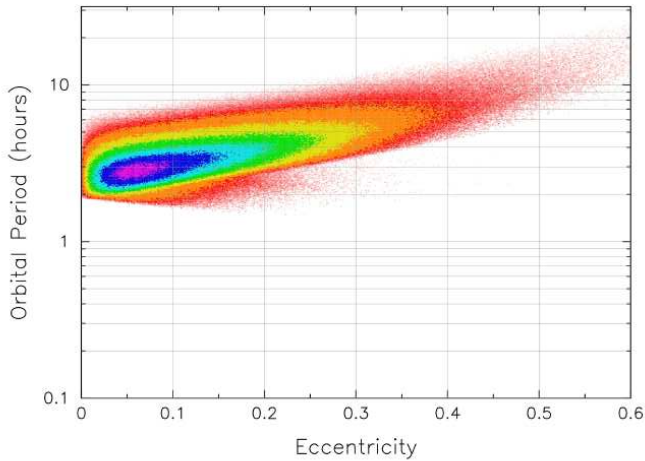,angle = -90,width =
    0.99\linewidth}}
\caption{Distribution of orbital periods and eccentricities of newly
formed BHRPs.  The colors were chosen according to the square root of
the number of systems that enter a given element of a $500\times 500$
array. In going from magenta to yellow, the number drops by a factor
of $\simeq$5.
\label{fig:peinit}}
\end{figure}

\section{Galactic Population}\label{sec:pop}

As the final step in our evolutionary study, we have combined the most
important elements of \S~3 into a code designed to generate a
population of newly formed BHRPs.  Several free parameters are chosen
via Monte Carlo methods from plausible distributions in such a way as
to illustrate the possible range of outcomes.  Fixed quantities in our
simulation include the maximum primary mass of $30\msun$ and the
common-envelope parameter $\gamma = 0.1$.  After a BHRP is formed, we
follow its further orbital evolution due to the emission of
gravitational radiation.  Our procedure is outlined below:

\medskip

\noindent
(a) The initial primary mass is chosen from $p(M_p^{\rm PB}) \propto
(M_p^{\rm PB})^{-2.5}$ over the range of $8\msun$ to $30\msun$.  The
upper limit is intermediate between scenarios I and II discussed in
\S~\ref{sec:pb}.

\medskip

\noindent
(b) The secondary mass, $M_s^{\rm PB}$, is drawn from a flat
distribution in mass ratios.  Systems with $M_s^{\rm PB}/M_p^{\rm PB}<
0.5$ are dropped since the mass transfer from the primary is likely to
be dynamically unstable (see \S~\ref{sec:stabmt}).

\medskip

\noindent
(c) The value of $\beta$ is chosen from the range 0.5--1 with uniform
probability.  We then compute the secondary mass, $M_s^{\rm MT}$,
after the phase of stable mass transfer from the primary.  We adopt
$25\msun$ as the mass threshold for black-hole formation from the
secondary, and cut systems with $M_s^{\rm MT} < 25\msun$.

\medskip

\noindent
(d) Neglecting possible correlations between $M_s^{\rm MT}$ and the
orbital parameters following the supernova, we assume $\sigma_k =
200\kms$ and draw $a_{\rm SN,cir}$ from the cumulative distribution
shown in Fig.~\ref{fig:circ}.  As in Fig.~\ref{fig:circ}, we assume
that the distribution vanishes at $a_{\rm SN,cir} = 20\au$.

\medskip

\noindent
(e) For a given secondary mass, we compute the mass and radius of the
core (see eqs.[\ref{eq:coremass}] and [\ref{eq:corerad}]) that is
exposed after the common-envelope phase.  If the the core overflows
its Roche lobe following the spiral-in, the system is cut.

\medskip

\noindent
(f) If a system survives the spiral-in, we assume that the secondary's
core loses 0--$2\msun$ in a stellar wind, chosen with uniform
probability.  The orbit is widened accordingly.

\medskip

\begin{figure}[t]
\centerline{\epsfig{file = 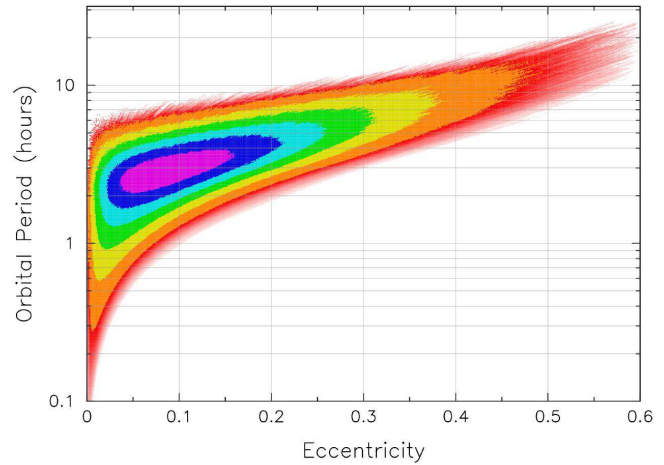,angle = -90,width =
    0.99\linewidth}}
\caption{Distribution of orbital periods and eccentricities of BHRPs
at the current epoch. The orbit of each binary has been evolved under
the action of gravitational radiation losses.  The colors are as in
Fig.~\ref{fig:peinit}.\label{fig:pece}}
\end{figure}

\noindent
(g) We consider both impulsive mass loss and a kick to the black hole
at the time of core collapse.  The fractional mass loss is chosen
uniformly from the range $0 < \Delta_{\rm BH} < 0.2$, and the kick is
drawn from a Maxwellian distribution with $\sigma_{k, {\rm BH}} =
200(M_{\rm NS}/M_{\rm BH})\kms$, so as to apply an average {\em
momentum} kick that is similar to neutron stars.

\medskip

\noindent
(h) We assume that BHRPs are formed at a constant rate and randomly
choose a birth time, $t_b$, from the range 0--10\,Gyr.

\medskip

\noindent
(i) After the formation of the BHRP, further orbital evolution is
driven by gravitational radiation, which we calculate using the
equations of \citet{peters64}.

\medskip

We followed $10^9$ binaries through the steps above.  For each
successful BHRP that is produced, the event is stored in a $500 \times
500$ array representing the distribution of initial orbital periods,
$P_b$, and eccentricities, $e$.  Figure~\ref{fig:peinit} shows this
distribution as a color image, where the colors were chosen according
to the square root of the number of binaries formed in each element of
the $P_b$--$e$ array.  In going from magenta to yellow, the number per
array element decreases by a factor of $\simeq$5.  Most of the newly
formed BHRPs have $2\hr < P_b < 7\hr$ and $e < 0.3$.  The effects of
supernova mass loss and black-hole kicks lead to a low-probability
tail that extends to $P_b \simeq 25\hr$ and $e \simeq 0.8$. Not
displayed is the distribution of black-hole masses, which ranges over
$M_{\rm BH} = 5$--$10\msun$, with a mean of $\simeq$$7\msun$.  Our
Monte Carlo approach allows us to directly estimate the birthrate,
since we are, in effect, explicitly calculating the survival fractions
listed in \S~\ref{sec:birth}.

\begin{deluxetable*}{l c c r@{.}l c r@{.}l  c c c}
\tabletypesize{\footnotesize}
\tablecaption{Results for Variations in the Standard Model}
\tablehead{
\colhead{Model}   &  \colhead{Description}  & \multicolumn{3}{r}{Birthrate ($10^{-7}\yr^{-1}$)} & 
\multicolumn{3}{c}{Number in Galaxy}  & \colhead{$\langle\tau_m({\rm Myr})\rangle$} 
& \colhead{$\langle e \rangle$} & \colhead{$\langle P_b({\rm hr})\rangle$}}
\startdata
1                 \dotfill  & Standard ($\gamma = 0.1$) & & 0&63 & & 5&4 & \phn85 & 0.078 & 3.5 \\
2\tablenotemark{a}\dotfill  & $\gamma = 0.07$ & & 0&03 & & 0&2 & \phn57 & 0.080 & 3.2 \\
3\tablenotemark{a}\dotfill  & $\gamma = 0.15$ & & 8&80 & & 103&0 & 118 & 0.078 & 3.8 \\
4\tablenotemark{a}\dotfill  & $\gamma = 0.30$ & & 110&00 & & 3800&0 & 345 & 0.079 & 5.2 \\
5\tablenotemark{b}\dotfill & $p(M_p^{\rm PB}) \propto (M_p^{\rm PB})^{-2}$ & & 0&54 & & 4&4 & \phn84 & 0.078 & 3.6 \\
6\tablenotemark{c}\dotfill & $p(q_{\rm PB}) \propto q_{\rm PB}^{1/2}$ & & 0&92 & & 7&6 & 101 & 0.079 & 3.7 \\
7\tablenotemark{d}\dotfill  & $(M_s^{\rm MT})_{\rm th} =20 M_\odot$ & & 1&50 & & 26&0 & 178 & 0.100 & 4.1 \\
8\tablenotemark{d}\dotfill  & $(M_s^{\rm MT})_{\rm th} =30 M_\odot$ & & 0&28 & & 1&3 & \phn47 & 0.069 & 3.3 \\
9\tablenotemark{e}\dotfill & $\sigma_k = 100\kms$ & & 4&20 & & 35&0 & \phn84 & 0.078 & 3.6 \\
10\tablenotemark{e}\dotfill & $\sigma_k = 300\kms$ & & 0&15 & & 1&3 & \phn85 & 0.079 & 3.6 \\
11\tablenotemark{f}\dotfill  & $1.25 R_c$ & & 0&14 & & 2&1 & 157 & 0.084 & 4.9 \\
12\tablenotemark{f}\dotfill & $1.5 R_c$ & & 0&02 & & 0&4 & 280 & 0.093 & 5.8 \\
13\tablenotemark{g}\dotfill  & $(\Delta M_{c,s})_{\rm wind} = 0$ & & 0&63 & & 2&5 & \phn40 & 0.072 & 3.0 \\
14\tablenotemark{h}\dotfill  & $\Delta_{\rm BH} =0 $ & & 0&63 & & 2&8 & \phn44 & 0.042 & 2.9 \\
15\tablenotemark{i}\dotfill & $\sigma_{k,{\rm BH}} = 0\kms$ & & 0&63 & & 4&8 & \phn76 & 0.058 & 3.6 \\
16\tablenotemark{i}\dotfill & $\sigma_{k,{\rm BH}} = 400(M_{\rm NS}/M_{\rm BH})\kms$ & & 0&63 & & 5&9 & \phn93 & 0.116 & 3.7 \\
\enddata
\tablenotetext{a}{Envelope binding-energy parameter associated with the 
spiral-in phase (eq.[\ref{eq:ce}]).}
\tablenotetext{b}{The initial mass function for the primary star.}
\tablenotetext{c}{The distribution of mass ratios for the primordial binaries.}
\tablenotetext{d}{Minimum post-transfer secondary mass that leaves a black-hole remnant.}
\tablenotetext{e}{The one-dimensional velocity dispersion for neutron-star kicks.}
\tablenotetext{f}{The radius of the secondary's core, nominally given by eq.(\ref{eq:corerad}), was 
arbitrarily increased by factors of 1.25 and 1.5 to allow for the 
effects of thermal bloating.}
\tablenotetext{g}{Maximum mass loss via a stellar wind from the exposed core of the secondary.}
\tablenotetext{h}{Fractional mass lost from the binary during black hole formation.}
\tablenotetext{i}{The one-dimensional velocity dispersion for black-hole kicks..}

\end{deluxetable*}

Given the initial $P_b$, $e$, and $M_{\rm BH}$, the orbit is evolved
forward in time using the equations of \citet{peters64}; $t_e$ is the
evolutionary time since the birth of the BHRP.  The evolution is
terminated when the system merges for $t_b + t_e < 10\gyr$, or when
$t_b + t_e = 10\gyr$ is reached.  For each BHRP that traverses the
$P_b$-$e$ plane, we store the amount of time spent in each element of
the $500 \times 500$ array; these times are accumulated for all the
systems.  The result is an array of total dwell times that represent
the probability density of finding a system at the current epoch with
a given $P_b$ and $e$.  Figure~\ref{fig:pece} displays the expected
current-epoch distribution of binary parameters.

Summing the dwell times over the entire array, dividing by the total
number of BHRPs formed successfully from $10^9$ primordial binaries,
and multiplying by the BHRP birthrate, we estimate the number of BHRPs
that currently inhabit the Galaxy.  In our standard model listed in
the steps above, we find $\calb \simeq 6\times 10^{-8}\yr^{-1}$ and
estimate that $\simeq$5 BHRPs populate the Galaxy.  This estimate can
also be obtained upon examination of Fig.~\ref{fig:pece}, where it is
evident that the average orbital period is 3-4\,hr, corresponding to a
short merger time of $\simeq$$(4$--$8)\times 10^7\yr$.

Table~1 gives an indication of how our results vary when we consider a
range of plausible input assumptions that differ from those in our
standard model.  For each quantity or function that is varied, all
else is left as in the standard model.  As already demonstrated, the
BHRP birthrate and total number in the Galaxy depend most strongly on
the common-envelope parameter $\gamma$.  In the very extreme case of
$\gamma = 0.3$, the birthrate and total number are
$\sim$$10^{-5}\yr^{-1}$ and $\simeq$4000, respectively.  For $\gamma =
0.15$, which is on the upper edge of plausibility, the total number is
$\simeq$100. The column labeled $\langle \tau_m({\rm Myr})\rangle$, an
average merger time, is defined to give the total number when
multiplied by the birthrate in column 3.  In columns 6 and 7, $\langle
e \rangle$ and $\langle P({\rm hr})\rangle$ are traditional means of
the distribution.


\section{Pulsar Detection and Timing}\label{sec:detect}

Our results suggest that at most $\simeq$100, but probably $\la$10,
BHRPs currently populate the Galactic disk.  We now consider what
fraction of these systems are detectable in radio pulsar surveys.  A
plausible first guess is that, as a class, BHRPs are similar
observationally to their DNS cousins.  A number of papers (e.g.,
Kalogera et al. 2001 and references therein) indicate that $<$1\% of
the total Galactic DNS population has been discovered among several
extensive pulsar surveys conducted over the past 30\,yr.  This
fraction incorporates the sensitivity and sky coverage of each survey,
the expected spatial and luminosity distributions of DNSs, and the
fraction ($\simeq$20--30\%) of pulsars beamed toward the Earth.  Such
a small detection efficiency is consistent with not yet having
discovered a BHRP.  If indeed there are $\simeq$100 BHRPs in the
Galaxy, then $\sim$1 system may be within reach of ongoing surveys
using the Parkes and Green Bank radio telescopes.  However, we view
this as being very optimistic, since 1\% is probably a firm upper
limit to the detection efficiency with current radio telescopes, and
because we predict a substantially smaller number of BHRPs.  
Realistically, there are likely to be important differences between
the BHRP and DNS populations that impact the BHRP detection
likelihood, three of which we now discuss in turn.

The recycling histories of BHRPs and DNSs may be distinct.  As
discussed in \S~\ref{sec:cesev}, it is possible that pulsars in BHRPs
have longer spin periods, $P_s$, and perhaps also larger surface
magnetic fields, $B$.  The spin-down luminosity scales as $B^2/P_s^4$.
Systematically longer spin periods may be compensated for by larger
magnetic fields, giving BHRPs luminosities similar to DNSs.

For reasons given in \S~\ref{sec:bhform}, the mean systemic speed
(e.g., $\simeq$$30\kms$) of newly formed BHRPs may be smaller than for
DNSs ($\ga$$100\kms$), resulting in a smaller scaleheight of BHRPs in
the Galactic disk: $\la$500\,pc for BHRPs, compared to $\ga$1\,kpc for
DNSs. Overall, we expect that the smaller scaleheight of BHRPs
somewhat increases their discovery probability relative to DNSs, but
not by a large factor.  It is also important to note that if BHRPs are
more concentrated toward the Galactic midplane, pulse smearing due to
dispersion in the interstellar medium, which decreases the
sensitivity, is more significant than for DNSs.  However, the modest
relative reduction in sensitivity is partly canceled if pulsars in
BHRPs have longer spin periods.

Perhaps the most robust difference between BHRPs and DNSs, in regard
to their detectability, is that the pulsars in BHRPs should, on
average, have larger orbital accelerations, because of their more
massive black-hole companions and generally short orbital periods.
Acceleration of the pulsar causes the measured pulse frequency to
drift over the integration time of the observation, thus smearing the
pulsar's signal in the Fourier power spectrum of the data and reducing
its amplitude \citep[e.g.,][]{anderson93,johnston91}.  For black-hole
masses in the range of 5--$10\msun$, the acceleration of a pulsar in a
BHRP is $\simeq$2 times that in a DNS with the same orbital period and
binary inclination, assuming that the orbits have small
eccentricities.  A search that explicitly probes for accelerated
pulsars may be {\em required} in order to detect a BHRP.  Such
acceleration searches are computationally expensive, but are now done
routinely.

Once a BHRP is discovered, the rewarding task of timing the pulsar
begins.  The initial Keplerian orbital parameters will be readily
determined with high precision.  Since BHRP orbits are probably very
compact, the effects of relativistic gravity will be quite pronounced.
Relativistic effects that have been measured in binary pulsars include
orbital precession, the combined effect of the second-order Doppler
shift and gravitational redshift (Einstein delay), the Shapiro
propagation delay, orbital contraction due to gravity-wave emission,
and the influence of geodetic precession on the measured pulse profile
\citep{damour92,weisberg02,stairs04b}.  Each of these effects is a
priori more significant in a BHRP than in a DNS, as are delays due to
relativistic aberration and light bending \citep[e.g.,][]{wex99}.  The
Keplerian parameters, coupled with the measurement of the Einstein or
Shaprio delay, will yield the mass of the black hole, which is
fundamentally important for studying models of black-hole formation,
in particular because the black holes in BHRPs have not accreted mass.
A BHRP also offers the tantalizing possibility of measuring
strong-field effects of relativistic gravity and providing stringent
constraints on theories alternative to Einstein's general relavity.
The high sensitivity of the planned Square Kilometer Array may be
required to tap into these more exotic problems (e.g., Kramer et
al. 2004 and references therein)---and perhaps to provide the first
BHRP detection.


\section{Conclusions}

We would like to underscore three main conclusions of our study.
(1)~There are major uncertainties in the formation of BHRPs, linked to
important details in the life and death of stars more massive than
$\simeq$$20\msun$, but mainly to a poor understanding of the
common-envelope phase.  Under quite optimistic circumstances, we
expect that the Galactic BHRP birthrate is $\sim$$10^{-7}\yr^{-1}$.
The birthrate may be much lower, and possibly {\em zero}.  The effects
of dramatic orbital shrinkage during the common-envelope phase,
prodigious stellar winds, and hypercritical accretion may preclude
BHRP formation altogether (see \S~\ref{sec:cesev}).  (2)~Plausible,
but possibly minimal, common-envelope shrinkage factors of
$\simeq$1000 lead to initial BHRP orbital periods of 2--10\,hr and an
average merger time of $\sim$$10^8\yr$.  (3)~The short merger times
and low birthrate imply that there are most likely no more than
$\sim$10 BHRPs currently in the Galactic disk, making the prospects
for their detection somewhat bleak.  As a direct comparison, we
estimate that there is probably no more than 1 BHRP in the Galaxy for
every 100--1000 double neutron stars, of which only 8 are currently
known.

Our simulations suggest that most BHRPs at the current epoch have
orbital periods of 1--6\,hr.  Due to supernova mass loss and possible
black-hole kicks, a typical eccentricity may be $\sim$0.1.  The
acceleration of the pulsar in such a binary is large, but does not
greatly hinder its detection given current computational resources.
If a BHRP is discovered, timing measurements would provide a wealth of
new constraints on theories of gravitation.


\acknowledgements

We would like to thank J. Dewi, R. Di Stefano, A. Loeb, and S. Ransom
for valuable discussions.  EP was supported by NASA and the Chandra
Postdoctoral Fellowship program through grant number PF2-30024.






\begin{thebibliography}{}

\bibitem[Alpar et al.(1982)]{alpar82} Alpar, M.~A., Cheng, A.~F.,
Ruderman, M.~A., \& Shaham, J.\ 1982, \nat, 300, 728

\bibitem[Anderson(1993)]{anderson93} Anderson, S. B. 1993,
Ph.D. thesis, Caltech

\bibitem[Arzoumanian et al.(1999)]{arzoumanian99} Arzoumanian, Z.,
Cordes, J.~M., \& Wasserman, I.\ 1999, \apj, 520, 696

\bibitem[Arzoumanian, Chernoff, \& Cordes(2002)]{arzoumanian02}
Arzoumanian, Z., Chernoff, D.~F., \& Cordes, J.~M.\ 2002, \apj, 568,
289

\bibitem[Belczynski et al.(2002)]{belkalbul02} Belczynski, K.,
Kalogera, V., \& Bulik, T.\ 2002, \apj, 572, 407

\bibitem[Bhattacharya \& van den Heuvel(1991)]{bhat91} Bhattacharya,
D.~\& van den Heuvel, E.~P.~J.\ 1991, \physrep, 203, 1

\bibitem[Brandt \& Podsiadlowski(1995)]{brandt95} Brandt, N.~\&
Podsiadlowski, P.\ 1995, \mnras, 274, 461

\bibitem[Brandt, Podsiadlowski, \& Sigurdsson(1995)]{brandt95b} 
Brandt, W.~N., Podsiadlowski, P., \& Sigurdsson, S.\ 1995, \mnras, 277, L35 

\bibitem[Brown(1995)]{brown95} Brown, G. E., 1995, \apj, 440, 270

\bibitem[Brown et al.(2001)]{brown01} Brown, G. E., Heger, A., Langer,
N., Lee, C.-H., Wellstein, S., \& Bethe, H. 2001, New Astronomy, 6,
457

\bibitem[Burgay et al.(2003)]{burgay03} Burgay, M., et al.\ 2003,
\nat, 426, 531

\bibitem[Cappellaro et al.(1999)]{cappellaro99} Cappellaro, E., Evans,
R., \& Turatto, M.\ 1999, \aap, 351, 459

\bibitem[Chevalier(1993)]{chevalier93} Chevalier, R.~A.\ 1993, \apjl,
411, L33

\bibitem[Chevalier(1996)]{chevalier96} Chevalier, R.~A.\ 1996, \apj,
459, 322

\bibitem[Chiosi \& Maeder(1986)]{chiosi86} Chiosi, C., \& Maeder, A.\
1986, \araa, 24, 329




\bibitem[Damour \& Taylor(1992)]{damour92} Damour, T., \& Taylor,
J.~H.\ 1992, \prd, 45, 1840



\bibitem[Dewi \& Tauris(2001)]{dewi01} Dewi, J.~D.~M.~\& Tauris,
T.~M.\ 2001, ASP Conf.~Ser.~229, Evolution of Binary and Multiple Star
Systems, ed. Ph. Podsiadlowski, S. Rappaport, A. R. King, F. D'Antona,
\& L. Burderi (San Francisco: ASP), 255


\bibitem[Dewi \& Pols(2003)]{dewi03} Dewi, J.~D.~M.~\& Pols, O.~R.\
2003, \mnras, 344, 629

\bibitem[Dymnikova(1986)]{dymnikova86} Dymnikova, I.~G.\ 1986, IAU
Symp. 114, Relativity in Celestial Mechanics and Astrometry: High
Precision Dynamical Theories and Observational Verifications,
ed. J. Kovalevsky \& V. A. Brumberg (Dordrecht: Reidel), 411

\bibitem[Eggleton(1983)]{eggleton83} Eggleton, P.~P.\ 1983, \apj, 268,
368

\bibitem[Faulkner et al.(2005)]{faulkner05} Faulkner, A.~J., et al.\
2005, \apjl, 618, L119


\bibitem[Fryer \& Kalogera(2001)]{fryer01} Fryer, C.~L.~\& Kalogera,
V.\ 2001, \apj, 554, 548

\bibitem[Goicoechea et al.(1992)]{goicoechea92} Goicoechea, L.~J.,
Mediavilla, E., Buitrago, J., \& Atrio, F.\ 1992, \mnras, 259, 281

\bibitem[Gourgoulhon \& Haensel(1993)]{gourgoulhon93} Gourgoulhon, E.,
\& Haensel, P. 1993, A\&A, 271, 187

\bibitem[Hansen \& Phinney(1997)]{hansen97} Hansen, B.~M.~S.~\&
Phinney, E.~S.\ 1997, \mnras, 291, 569

\bibitem[Hulse \& Taylor(1975)]{hulse75} Hulse, R.~A.~\& Taylor,
J.~H.\ 1975, \apjl, 195, L51

\bibitem[Humphreys(1984)]{humphreys84} Humphreys, R. M. 1984, in
Observational Test of Stellar Evolution Theory, ed. A. Maeder \&
A. Renzini (Dordrecht, Reidel), 279

\bibitem[Hurley, Pols, \& Tout(2000)]{hurley00} Hurley, J.~R., Pols,
O.~R., \& Tout, C.~A.\ 2000, \mnras, 315, 543

\bibitem[Ivanova et al.(2003)]{ivanova03} Ivanova, N., Belczynski, K.,
Kalogera, V., Rasio, F.~A., \& Taam, R.~E.\ 2003, \apj, 592, 475

\bibitem[Johnston \& Kulkarni(1991)]{johnston91} Johnston, H.~M.~\&
Kulkarni, S.~R.\ 1991, \apj, 368, 504

\bibitem[Jonker \& Nelemans(2004)]{jonker04} Jonker, P. G., \&
Nelemans, G. 2004, MNRAS, 354, 355

\bibitem[Joss \& Rappaport(1983)]{joss83} Joss, P.~C.~\& Rappaport,
S.~A.\ 1983, \nat, 304, 419

\bibitem[Kalogera \& Webbink(1998)]{kalogera98} Kalogera, V.~\&
Webbink, R.~F.\ 1998, \apj, 493, 351

\bibitem[Kalogera et al.(2001)]{kalogera01} Kalogera, V., Narayan, R.,
Spergel, D.~N., \& Taylor, J.~H.\ 2001, \apj, 556, 340

\bibitem[Kalogera et al.(2004)]{kalogera04} Kalogera, V., et al.\
2004, \apjl, 601, L179

\bibitem[Kramer et al.(2004)]{kramer04} Kramer, M., Backer, D.~C.,
Cordes, J.~M., Lazio, T.~J.~W., Stappers, B.~W., \& Johnston, S.\
2004, New Astronomy Review, 48, 993

\bibitem[Laguna \& Wolszczan(1997)]{laguna97} Laguna, P.~\& Wolszczan,
A.\ 1997, \apjl, 486, L27

\bibitem[Langer \& Maeder(1995)]{langer95} Langer N., \& Maeder A.,
1995, A\&A, 295, 685

\bibitem[Lipunov et al.(1994)]{lipunov94} Lipunov, V.~M., Postnov,
K.~A., Prokhorov, M.~E., \& Osminkin, E.~Y.\ 1994, \apjl, 423, L121

\bibitem[Lyne et al.(2004)]{lyne04} Lyne, A.~G., et al.\ 2004,
Science, 303, 1153


\bibitem[Narayan, Piran, \& Shemi(1991)]{narayan91} Narayan, R., 
Piran, T., \& Shemi, A.\ 1991, \apjl, 379, L17 

\bibitem[Peters(1964)]{peters64} Peters, P.~C.\ 1964, Physical Review,
136, 1224

\bibitem[Pfahl, Rappaport, \& Podsiadlowski(2002)]{pfahl02a} Pfahl,
E., Rappaport, S., \& Podsiadlowski, P.\ 2002, \apj, 573, 283 (PRP02)

\bibitem[Pfahl et al.(2002)]{pfahl02b} Pfahl, E., Rappaport, S.,
Podsiadlowski, P., \& Spruit, H.\ 2002, \apj, 574, 364

\bibitem[Podsiadlowski et al.(2002)]{podsi02} Podsiadlowski, Ph.,
Nomoto, K., Maeda, K., Nakamura, T., Mazzali, P., \& Schmidt, B. 2002,
ApJ, 567, 491

\bibitem[Podsiadlowski, Rappaport, \& Han(2003)]{podsi03}
Podsiadlowski, P., Rappaport, S., \& Han, Z.\ 2003, \mnras, 341, 385
(PRH03)

\bibitem[Podsiadlowski et al.(2004)]{podsi04} Podsiadlowski, Ph.,
Langer, N., Poelarends, A., Rappaport, S., Heger, A., \& Pfahl,
E. 2004, ApJ, 612, 104

\bibitem[Portegies Zwart, Verbunt, \& Ergma(1997)]{zwart97} Portegies
Zwart, S.~F., Verbunt, F., \& Ergma, E.\ 1997, \aap, 321, 207


\bibitem[Ransom et al.(2004)]{ransom04} Ransom, S.~M., et al. \ 2004,
\apjl, 609, L71

\bibitem[Sigurdsson(2003)]{sigurdsson03} Sigurdsson, S.\ 2003, in ASP
Conf. Ser. 302, Radio Pulsars, ed. M. Bailes, D. J. Nice, \&
S. E. Thorsett (San Francisco: ASP), 391

\bibitem[Sipior \& Sigurdsson(2002)]{sipior02} Sipior, M.~S., \&
Sigurdsson, S.\ 2002, \apj, 572, 962

\bibitem[Sipior, Portegies Zwart, \& Nelemans(2004)]{sipior04} Sipior,
M. S., Portegies Zwart, S., \& Nelemans, G. 2004, MNRAS, submitted,
astro-ph/0407268

\bibitem[Smarr \& Blandford(1976)]{smarr76} Smarr, L.~L.~\& Blandford,
R.\ 1976, \apj, 207, 574


\bibitem[Stairs(2004)]{stairs04} Stairs, I.~H.\ 2004, Science, 304,
547

\bibitem[Stairs et al.(2004)]{stairs04b} Stairs, I.~H., Thorsett,
S.~E., \& Arzoumanian, Z.\ 2004, Phys. Rev. Lett., 93, 141101

\bibitem[Tauris \& van den Heuvel(2003)]{tauris03} Tauris, T. M., \&
van den Heuvel, E. P. J. \ 2003, to appear in ``Compact Stellar X-ray
Sources, eds. W. H. G. Lewin \& M. van der Klis, astro-ph/0303456

\bibitem[Voss \& Tauris(2003)]{voss03} Voss, R.~\& Tauris, 
T.~M.\ 2003, \mnras, 342, 1169 

\bibitem[Webbink(1984)]{webbink84} Webbink, R.~F.\ 1984, \apj, 277,
355

\bibitem[Weisberg \& Taylor(2002)]{weisberg02} Weisberg, J.~M., \&
Taylor, J.~H.\ 2002, \apj, 576, 942

\bibitem[Wellstein \& Langer(1999)]{wellstein99} Wellstein, S.~\&
Langer, N.\ 1999, \aap, 350, 148

\bibitem[Wex \& Kopeikin(1999)]{wex99} Wex, N.~\& Kopeikin, 
S.~M.\ 1999, \apj, 514, 388 

\bibitem[Woosley \& Weaver(1995)]{woosley95} Woosely, S. E., Weaver,
T. A. 1995, ApJS, 101, 181

\end{thebibliography}
\end{document}